\documentstyle[prd,aps,floats]{revtex}

\begin{document}

\draft
\input epsf
\twocolumn[\hsize\textwidth\columnwidth\hsize\csname
@twocolumnfalse\endcsname

\title{Topological Defects Formation after Inflation 
                                   on Lattice Simulation} 

\author{S. Kasuya and M. Kawasaki}
\address{Institute for Cosmic Ray Research, University of Tokyo,
  Tanashi, Tokyo 188-8502, Japan}

\date{April 28, 1998}

\maketitle

\begin{abstract}
We consider the formation of topological defects after inflation. 
In order to take into account the effects of the rescattering of
fluctuations, we integrate the classical equation that describes the
evolution of a complex scalar field on the two-dimensional lattice with
a slab symmetry. The growth of fluctuations during preheating 
is found not to be enough for defect formation, and rather a long
stage of the rescattering of fluctuations
after preheating is necessary. We conclude that the topological
defects are not formed if the breaking scale $\eta$ is lager than 
$ \sim (2 - 3)\times 10^{16} {\rm GeV}$.
\end{abstract}

\pacs{PACS numbers: 98.80.Cq, 11.27.+d
      \hspace{5cm} hep-ph/9804429}

%\newpage
\vskip2pc]

\setcounter{footnote}{1}
\renewcommand{\thefootnote}{\fnsymbol{footnote}}

%%%%%%%%%%%%%%%%%%%%%%%%%%%%%%%%%%%%%%%%%%%%%%%%%%%%%%%%%%%%%%%%%
\section{Introduction}
%%%%%%%%%%%%%%%%%%%%%%%%%%%%%%%%%%%%%%%%%%%%%%%%%%%%%%%%%%%%%%%%%%
Inflation \cite{Guth,Sato} was invented in order to
overcome several problems in the 
standard hot big bang universe, such as the flatness and  horizon
problems. Moreover, harmful topological defects created before 
inflation 
were diluted away, and are almost absent in the present universe. If
some topological defects such as monopoles or domain walls are
produced after inflation, they are disastrous since they will soon
dominate the energy density of the universe. Therefore, the reheating
temperature cannot be as high 
as the grand unified theory (GUT) scale so as not to produce the GUT
monopoles. 

It was recently recognized that topological defects may be formed
during preheating \cite{KLS2,Tkachev,KK2}. Preheating is the very
beginning of the reheating process, and occurs because of the
parametric
resonant effects \cite{KLS1,Shtanov,Boyan1,Yoshi}. Its essence is
an {\it induced} effect in the sense that the presence of the produced
particles stimulates further decay of the coherently oscillating
scalar field into those particles \cite{KK1}. This phenomenon is thus
peculiar to bosonic particles that obey the Bose-Einstein
statistics.  

In the preheating stage, very large non-thermal
fluctuations are produced, 
$\langle \delta \phi^2 \rangle \sim c^2 M_p^2$ where $M_p$ is the
Planck mass and $c=10^{-2}-10^{-3}$ \cite{KLS2,Tkachev}. 
These fluctuations change
the shape of the effective potential of the field $\phi$ to restore
its symmetry if the potential $V(\phi)$ is of spontaneous
symmetry-breaking type. Later, when
the amplitude of these fluctuations is redshifted away by the cosmic
expansion, the symmetry is spontaneously broken and topological
defects may be created. Thus, the mechanism 
for producing the topological defects is somewhat
similar to the Kibble mechanism in high temperature
theory. However, we showed in Ref.\cite{KK2} that the amplitude of the 
fluctuations does not grow larger than that of the homogeneous mode
for $V(\phi)=\lambda(\phi^2-\eta^2)^2$, where $\lambda$ is the
self-coupling constant. It implies that 
the symmetry restoration does not take place and hence the 
topological defects are not produced. Therefore, we considered
that topological defects are formed through another
mechanism which is based on the following two facts. The first one is
that there are two minima of the effective 
potential in the radial direction in the simple U(1) theory. The
second is that some fluctuations exist initially at the preheating
stage. They are produced during inflation and stretched far beyond the 
horizon to become classical fields. Therefore, the initial condition
for the homogeneous field in some region in the universe is different
from that in another region in the universe. The dynamics of each
region in the universe is thus different and the final value of the
field (the minimum into which the field settles down) is different,
which leads to the defect formation. 

However, in Ref.~\cite{KK2}, we used the Hartree approximation so that
the interactions between particles with different momenta are not
fully included. As we will see later, rescatterings 
\cite{KT1,KT2,PR,KLS3,GKLS} are quite important to determine the 
dynamics of the defect formation as suggested by 
Refs.~\cite{Kofman1,Kofman2}. To deal with the effect of the
rescattering, we must go on to the lattice simulation
\cite{KT1,KT2,PR}. Therefore, in this paper, we integrated the
equations describing the dynamics in a real space on the
lattice. There are some difficulties in the simulations on
lattice. One of the difficulties is how we can implement the quantum
fluctuations of the fields on the lattice. We rely on the
idea of Ref.\cite{KT1} and do not discuss in detail (actually,
it is not sensitive to the initial conditions for the fluctuations if
the order of magnitude is properly taken \cite{PR}). The other one
is the box size and the lattice size. The lattice size must be small 
enough for identifying the defects. The box size should not be taken
too much smaller than the horizon size at the end of the
simulation, otherwise we miss the defects whose density is 
$O(10)$ per horizon. We also take the proper lattice and box
sizes so that we can see the resonance effects. From these
requirements, we calculate the evolution of a complex scalar field in
a two-dimensional lattice abandoning to calculate it in three dimensions
because of the lack of machine memories. 

\begin{figure}[t]
\centering
\hspace*{-7mm}
\leavevmode\epsfysize=8cm \epsfbox{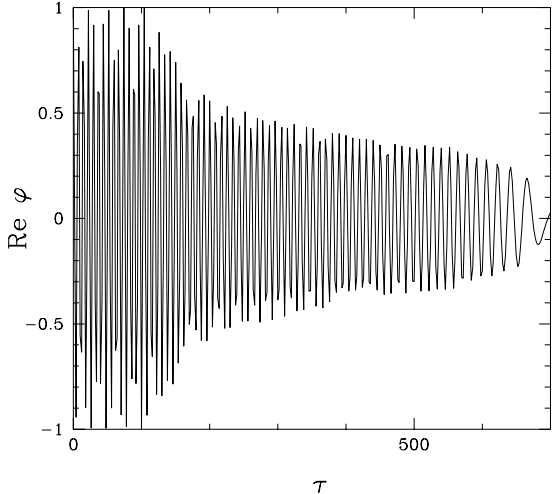}\\[2mm]
\caption[fig-ho]{\label{fig-ho} 
Evolution of the real part of the homogeneous field for 
$\eta=10^{16} {\rm GeV} $. }
\end{figure}

\begin{figure}[t]
\centering
\hspace*{-7mm}
\leavevmode\epsfysize=8cm \epsfbox{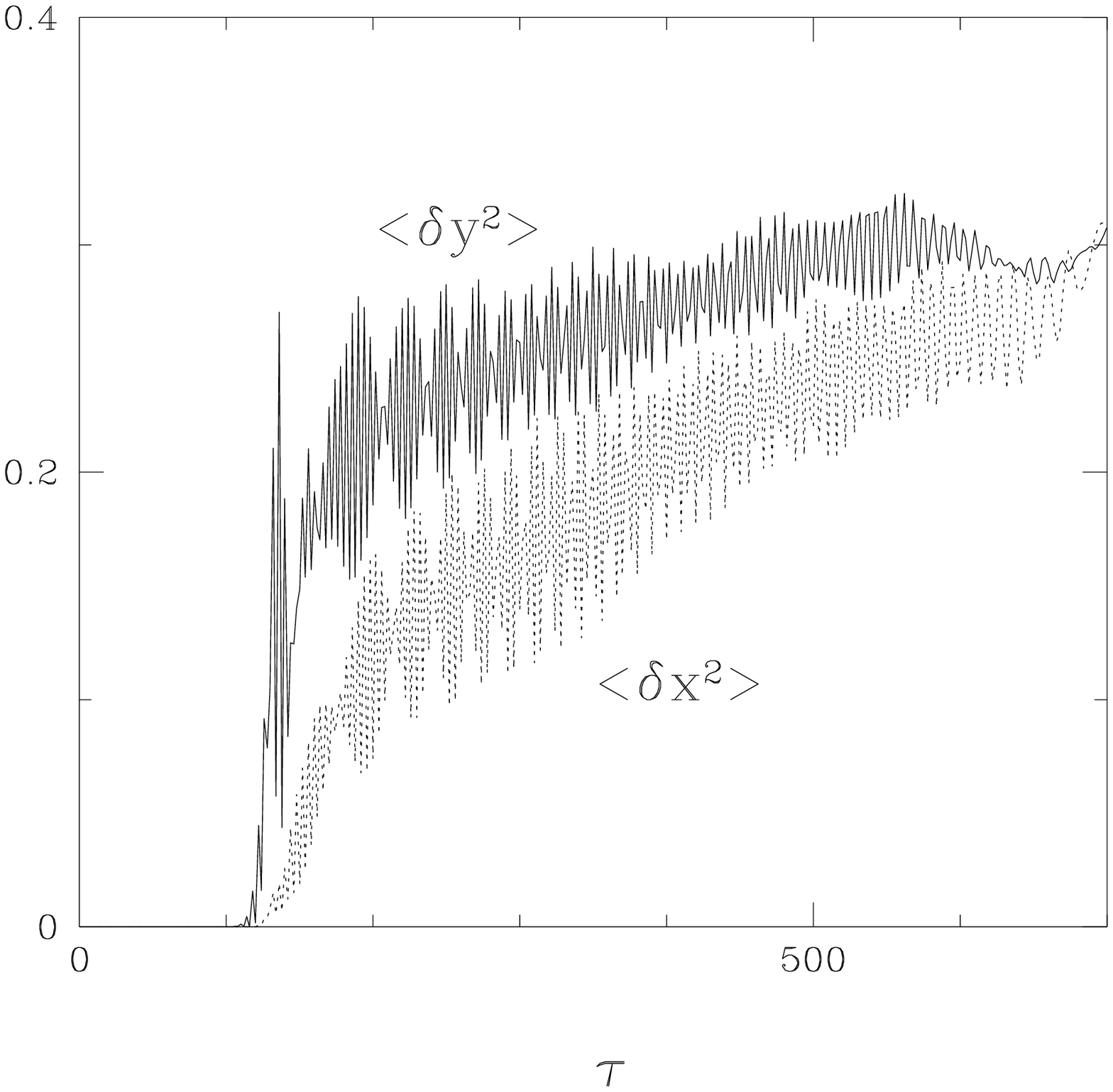}\\[2mm]
\caption[fig-fl-1]{\label{fig-fl-1} 
Evolution of the fluctuations for $\eta=10^{16} {\rm GeV}$. }
\end{figure}

\begin{figure}[t]
\centering
\hspace*{-7mm}
\leavevmode\epsfysize=8cm \epsfbox{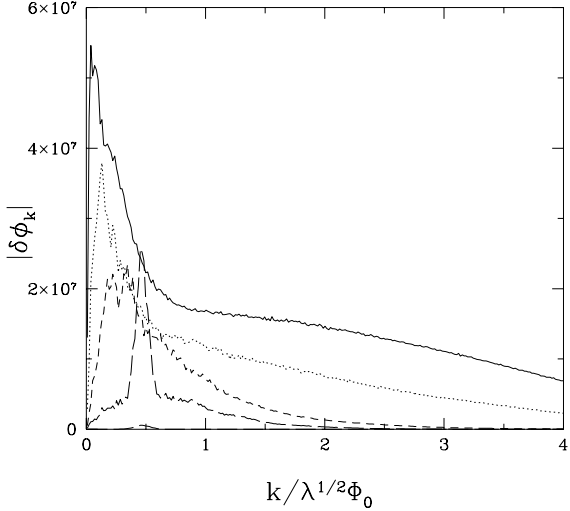}\\[2mm]
\caption[fig-spec-1]{\label{fig-spec-1} 
Spectrum of the fluctuations for $\eta=10^{16} {\rm GeV}$. These lines 
shows $\tau=800,500,200,130,100$ from top to bottom.}
\end{figure}

\begin{figure}[t]
\centering
\hspace*{-7mm}
\leavevmode\epsfysize=8cm \epsfbox{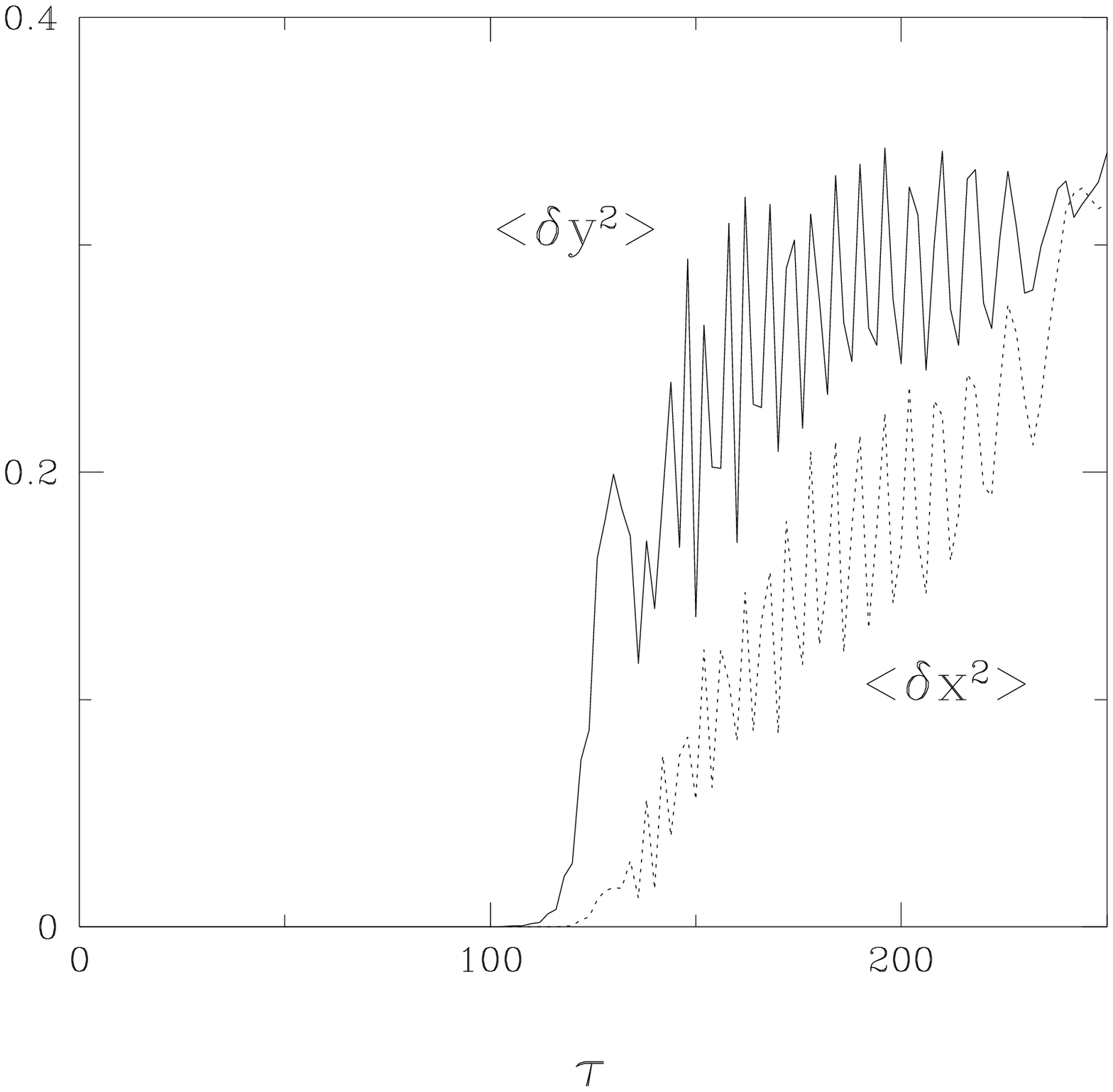}\\[2mm]
\caption[fig-fl-2]{\label{fig-fl-2} 
Evolution of the fluctuations for $\eta=3\times 10^{16} {\rm GeV}$. }
\end{figure}

\begin{figure}[t]
\centering
\hspace*{-7mm}
\leavevmode\epsfysize=8cm \epsfbox{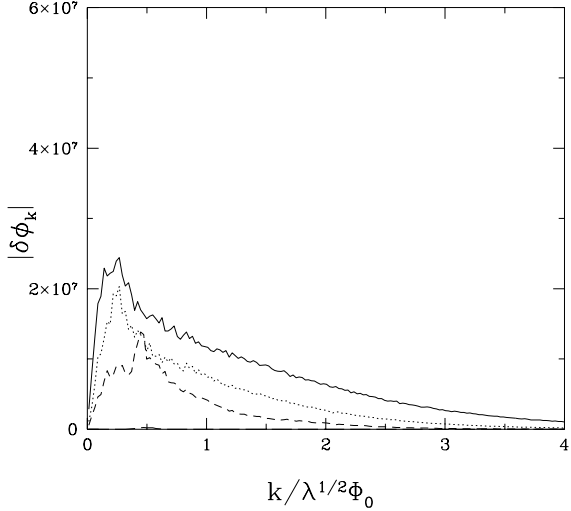}\\[2mm]
\caption[fig-spec-2]{\label{fig-spec-2} 
Spectrum of fluctuations for $\eta=3\times 10^{16} {\rm GeV}$. These
lines shows $\tau=250,200,150,100$ from top to bottom.}
\end{figure}

\begin{figure}[t]
\centering
\hspace*{-7mm}
\leavevmode\epsfysize=8cm \epsfbox{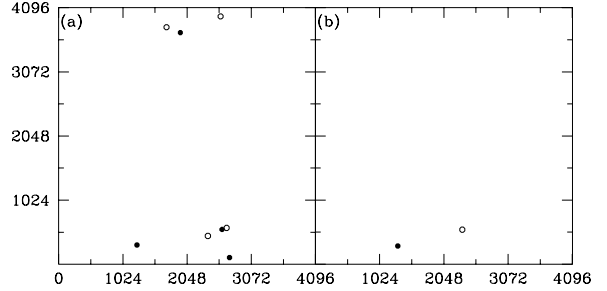}\\[2mm]
\caption[fig-r-def]{\label{fig-r-def} 
Evolution of defects for $\eta=2\times 10^{16} {\rm GeV}$ on lattice
at (a) $\tau=1000$ and (b) $\tau=1200$.
Dots and circles denote defects and anti-defects, respectively.}
\end{figure}

\begin{figure}[t]
\centering
\hspace*{-7mm}
\leavevmode\epsfysize=8cm \epsfbox{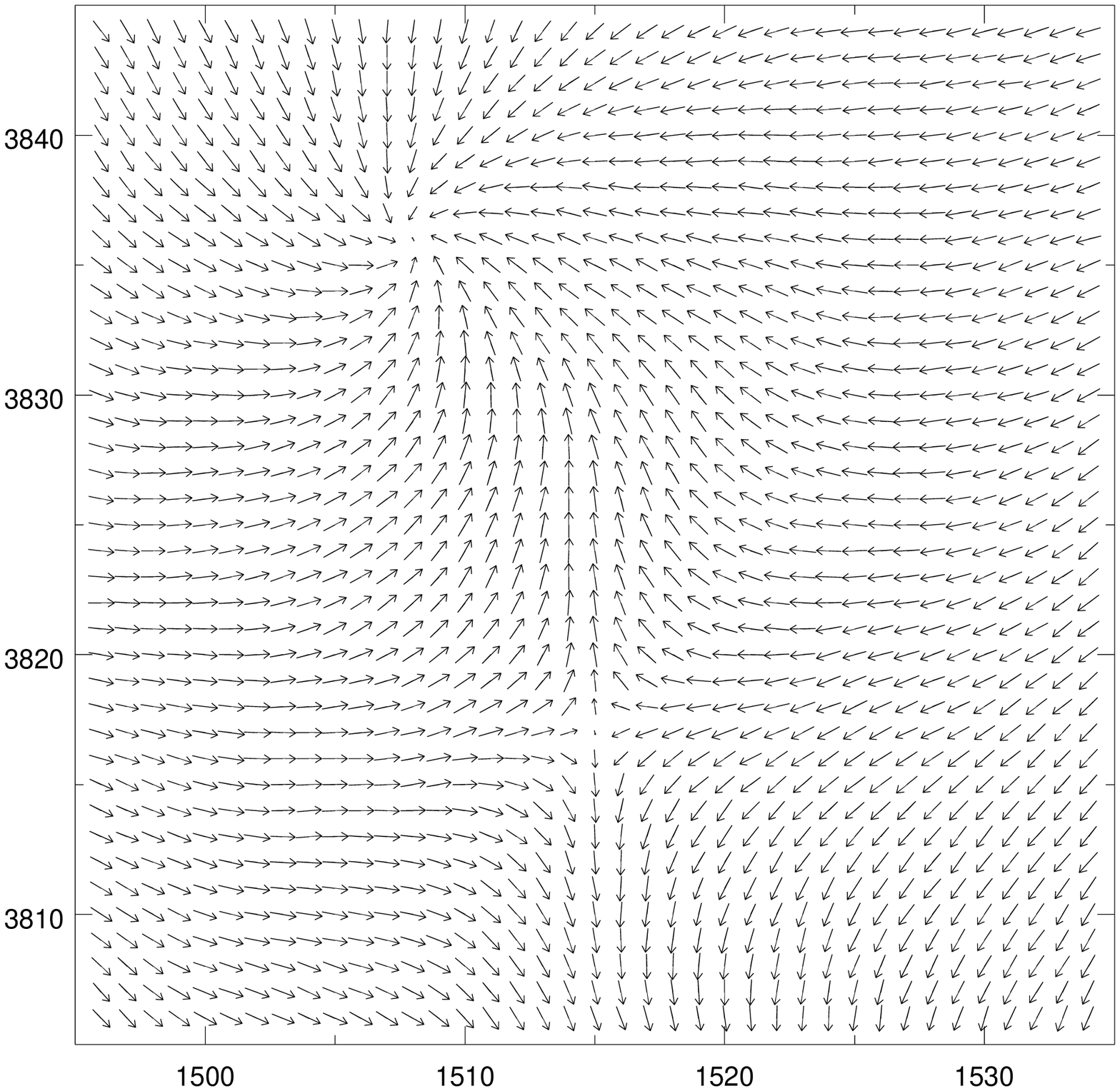}\\[2mm]
\caption[fig-phase]{\label{fig-phase} 
Phase of the field around the string and anti-string at $x \sim 1510$
and $y \sim 3830$ in Fig.~\ref{fig-p15-def}(d). }
\end{figure}

\begin{figure}[t]
\centering
\hspace*{-7mm}
\leavevmode\epsfysize=8cm \epsfbox{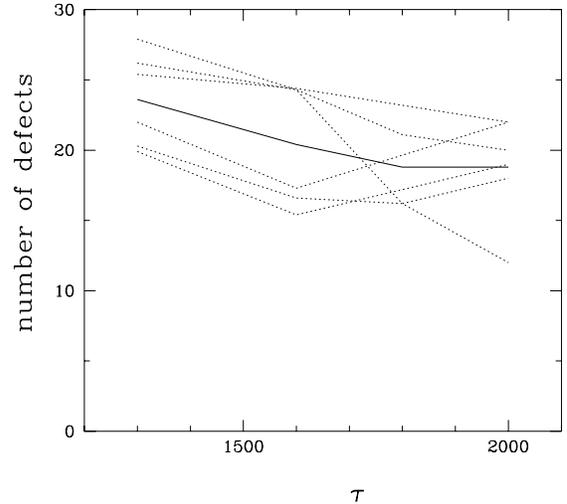}\\[2mm]
\caption[fig-p1-scale]{\label{fig-p1-scale} 
Changes of the number of the defects for $\eta=10^{16} {\rm GeV}$ with 
various initial conditions of fluctuations. The solid line denotes
the average.} 
\end{figure}

\begin{figure}[t]
\centering
\hspace*{-7mm}
\leavevmode\epsfysize=8cm \epsfbox{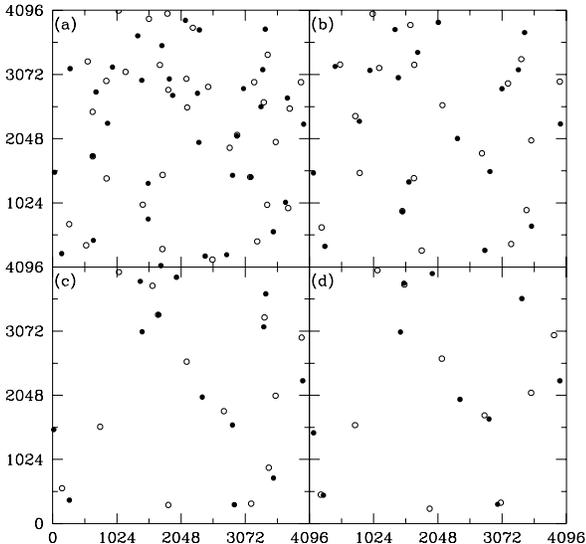}\\[2mm]
\caption[fig-p15-def]{\label{fig-p15-def} 
Evolution of the defects for $\eta=10^{16} {\rm GeV}$ on the lattice
at (a) $\tau=1300$, (b) $\tau=1600$, (c) $\tau=1800$, and 
(d) $\tau=2000$. Dots and circles denote defects and anti-defects,
respectively.}
\end{figure}

\begin{figure}[t]
\centering
\hspace*{-7mm}
\leavevmode\epsfysize=8cm \epsfbox{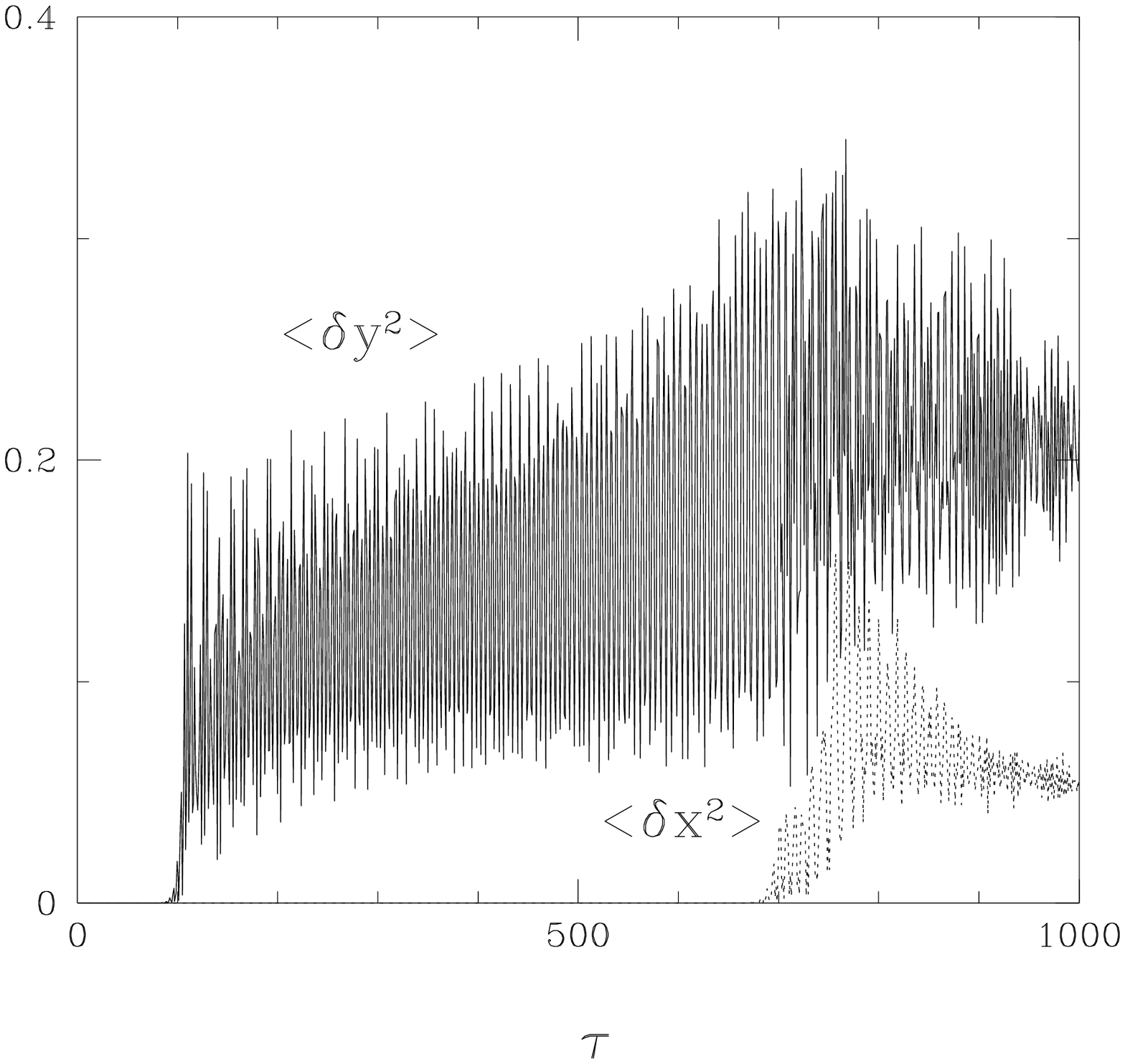}\\[2mm]
\caption[fig-fl-3]{\label{fig-fl-3} 
Evolution of the fluctuations for $\eta=10^{16} {\rm GeV}$ using the 
Hartree approximation in two dimensions in space. }
\end{figure}

%%%%%%%%%%%%%%%%%%%%%%%%%%%%%%%%
\section{Model}
%%%%%%%%%%%%%%%%%%%%%%%%%%%%%%%%
Let us consider the wine bottle potential
\begin{equation}
    V(\Phi) = \frac{\lambda}{2}(|\Phi|^2 - \eta^2)^2,
\end{equation}
where $\Phi$ is the complex scalar field, $\lambda$ is the very small
self coupling constant, and $\eta$ is the breaking scale. This model
has a global U(1) symmetry, and cosmic strings are formed when the
symmetry is spontaneously broken. The equation of motion becomes
\begin{equation}
  \label{eom-1}
    \ddot{\Phi} + 3H\dot{\Phi} - \frac{1}{a^2}\nabla^2\Phi 
                               + \lambda(|\Phi|^2 - \eta^2)\Phi = 0,
\end{equation}
where $H$ is the Hubble parameter and the dot denotes differentiation
with respect to time $t$. Rescaling as
\begin{eqnarray}
  \label{re-sf}
    a(\tau)d\tau & = & \sqrt{\lambda}\Phi_0 a(0)dt, \\
    \varphi & = & \frac{\Phi a(\tau)}{\Phi_0 a(0)}, \\
    \xi & = & \sqrt{\lambda}\Phi_0 a(0)x,
\end{eqnarray}
where $\Phi_0\equiv|\Phi(0)|$, then, setting $a(0)=1$, 
Eq.(\ref{eom-1}) becomes 
\begin{equation}
  \label{eom-2}
    \varphi^{\prime\prime} - \frac{a^{\prime\prime}}{a}\varphi
             - \nabla^2_{\xi}\varphi 
             + ( |\varphi|^2 - \tilde{\eta}^2 a^2 )\varphi =0,
\end{equation}
where $\tilde{\eta}\equiv\eta/\Phi_0$ and the prime denotes
differentiation with respect to $\tau$. The second term of LHS can be
omitted since the energy density of the universe behaves like
radiation at the early time and also the scale factor becomes very
large later. Therefore, we assume that the universe is radiation
dominated. In this case, the rescaled Hubble parameter, 
$h(\tau)\equiv H(\tau)/\sqrt{\lambda}\Phi_0$, and the scale factor
$a(\tau)$ become \cite{KK2}
\begin{equation}
    h(\tau) = \frac{\sqrt{2}}{3}a^{-2}(\tau),
\end{equation}
and
\begin{equation}
  \label{sf}
    a(\tau) = \frac{\sqrt{2}}{3}\tau + 1,
\end{equation}
respectively, when $\Phi$ is assumed to be an inflaton (even if $\Phi$ 
is not an inflaton, results are the same as in the case of rescaling
the breaking scale in an appropiate way, see Ref.~\cite{KK2}) 
and we put $a(0)=1$. 
Since the physical length and the horizon grow proportional to $a$ and
$a^2$, respectively, the rescaled horizon grows proportional to
$a$. The initial length of 
the horizon is $\ell_h(0)=3/\sqrt{2} \approx 2.12$. Therefore, the box
size should be larger than the horizon size at the end of calculation, 
$\ell_h(\tau)=\ell_h(0)a(\tau)$. 

The width of the topological defect is $(\sqrt{\lambda}\eta)^{-1}$
which corresponds to $(\tilde{\eta}a(\tau))^{-1}$ in the rescaled
scale. Since it decreases with time, one lattice length should be at
least comparable with the defect width at the end of the calculation. 

Furthermore, in order to see the resonance effects, both the lattice
size should be small enough and the box size should be large enough. 
Typical resonant momentum is $k \simeq \sqrt{\lambda}\Phi_0$ 
at the beginning. Since, in the $\lambda|\Phi|^4$ theory, the
resonance band does not change because of cosmic expansion, the lattice
size should be smaller than 
$\Delta x \sim k_{res}^{-1} \sim (\sqrt{\lambda}\Phi_0)^{-1}$. It
corresponds to $\sim 1$ in a rescaled unit. In addition, it is
necessary to have the box size large enough for a good resolution in
the resonance band. 

Three requirements lead us to take very large lattice sizes at the 
expense of one dimension in space. Thus we integrate Eq.(\ref{eom-2})
on a $4096 \times 4096$ lattice with a slab symmetry in a $z$-direction. 
Therefore, the defects that we see on the lattice are the
cosmic strings stretched infinitely along the $z$-direction.
  
For the initial conditions we take 
\begin{eqnarray}
    \varphi_x \equiv {\rm Re}\varphi(0) & = &
                               1 + \delta\varphi_x({\bf x}), \\
    \varphi_y \equiv {\rm Im}\varphi(0) & = &
                               \delta\varphi_y({\bf x}),         
\end{eqnarray}
where the homogeneous part comes from its definition (we call
$x$-direction for real direction and $y$ for imaginary), 
and $\delta\varphi_{x,y}(x)$ is a  small random variable representing
the fluctuations (we owe the fact that the semiclassical description
can be used to Ref.~\cite{KT1}). We also put small random values for
velocities. We integrated Eq.(\ref{eom-2}) until the time when the
horizon grew as large as the  
box size for five different breaking scales, $\eta=10^{16}, 2\times 
10^{16}, 3\times 10^{16}, 5\times 10^{16}$, and $10^{17} {\rm GeV}$. 
For example, we integrate until $\tau_{max}=2000$ for 
$\eta=10^{16} {\rm GeV}$. To check the accuracy of our code, we
calculate the energy density of the field for the  
$V(\Phi)=\lambda|\Phi|^4$ model, and it conserves within $0.36\%$
during preheating and $0.07\%$ after. (Notice that the energy density 
is not constant in the wine bottle type potential.)
We identify the defect if there is a winding number at that point. 

%%%%%%%%%%%%%%%%%%%%%%%%%%%%%
\section{Results}
%%%%%%%%%%%%%%%%%%%%%%%%%%%%%
We will first show the general feature of the dynamics taking 
$\eta = 10^{16} {\rm GeV}$ as an example. We take the lattice size
$\Delta\xi=0.5$ and 
$\langle \delta\varphi^2({\bf x}) \rangle^{1/2} = 10^{-7}-10^{-10}$ in
this case. Figure~\ref{fig-ho} shows the evolution of the
real part of the homogeneous field. Preheating takes
place at $\tau \sim 100$ and ends at $\tau \sim 150$. As can be seen
in Fig.~\ref{fig-fl-1}, the fluctuation in $y$-direction grows first and
the $x$-direction follows it. This is because the field feels no
potential 
in the phase direction, which means that the $y$-direction corresponds 
to the Nambu-Goldstone mode. During preheating, only those modes in
the resonance band grow exponentially through the parametric resonance 
effect (see Fig.~\ref{fig-spec-1}). Since the mode equations of the
fluctuations in $x$- and $y$-directions can be written as
\begin{eqnarray}
    \delta x_k^{\prime\prime} 
         + [ \tilde{k}_x^2 + 3x^2 
                 - \tilde{\eta}^2a^2 ] \delta x_k & = & 0, \\
   \label{y-eom}
     \delta y_k^{\prime\prime} 
         + [ \tilde{k}_y^2 + x^2 
                 - \tilde{\eta}^2a^2 ] \delta y_k & = & 0,
\end{eqnarray}
where back reactions are neglected and
$\tilde{k}=k/\sqrt{\lambda}\Phi_0$, the resonance mode for each
direction is 
$3/2 < \tilde{k}_x^2 < \sqrt{3}$ and $\tilde{k}_y^2 < 1/2$. As we
mentioned above, the fluctuation in $y$-direction grows faster. In the
actual calculation, the fastest growing mode is found to be 
$k \simeq 0.47$,
which coincides with the initial horizon size, 
$k^{-1} \simeq \ell_h(0)$. 
We also change the initial amplitude
of fluctuations. The difference only appears in the time at the end of 
preheating $\tau_{ph}$ but its change does not shorten the duration
of the rescattering so much [see Eq.(\ref{tau-ph}) below].  

After the exponential growth of the fluctuations (the preheating
stage), there is a rather long stage of oscillations of the
homogeneous mode with small amplitude, 
which ends when the field settles down into the minimum of the
potential at $\tau \sim 700$. During this stage, rescatterings of the
fluctuations become important. We can see that the amplitude of the 
fluctuations slightly grows in Fig.~\ref{fig-fl-1}, but redistribution 
of the spectrum is much more efficient. 

Let us consider the criterion for the defect formation. The most
simple idea is to compare the time when the amplitude of fluctuations
grows because 
of the parametric resonance ($\tau_{ph}$) with that when the field
settles down into the minimum of its potential ($\tau_{fall}$). 
More precisely, $\tau_{ph}$ is the time when the homogeneous field
decays through the parametric resonance until the back reactions of
the created fluctuations cannot be neglected. This usually happens
when the amplitude of fluctuations becomes as large as that of the
homogeneous mode. It is estimated from
\begin{equation}
  \label{tau-ph}
    \delta \varphi(0) \exp(\mu \tau_{ph}) \sim 1,
\end{equation}
where $\mu$ is the effective growth exponent. In the case of 
Eq.(\ref{y-eom}), the maximum value of $\mu$ is 
$\mu_{max} \simeq 0.147$ \cite{Boyan2,GKLS}. 
Since $\tau_{ph} \simeq 135$ for 
$\delta \varphi(0)=10^{-7}$ which we take in Fig.~\ref{fig-fl-1},
$\tau_{ph} \simeq 150$ for $\delta \varphi(0)=10^{-8}$, and
$\tau_{ph} \simeq 180$ for $\delta \varphi(0)=10^{-10}$ in our
calculations, we get $\mu_{eff} \sim 0.12$.~\footnote{%%
$\mu_{eff}$ is estimated from Eq.(\ref{tau-ph}). Since the resonance
does not occur from the very begining and becomes effective a little
later, the more conventional estimation is $\mu \simeq
0.141$. Actually, the square of amplitude of the fluctuation grows
from $\simeq 3\times10^{-15}$ to $\simeq 0.1$ during resonance 
($\Delta\tau = 110$) for $\delta \varphi(0) = O(10^{-7})$.}  
On the other hand, the amplitude of the homogeneous field decreases
due to cosmic expansion, and at $\tau = \tau_{fall}$,
\begin{equation}
    \frac{a(\tau_{fall})}{a(0)} \sim \frac{\Phi_0}{\eta}.
\end{equation}
Thus $\tau_{ph} \gtrsim \tau_{fall}$ for 
$\eta \gtrsim 6.7 \times 10^{16} {\rm GeV}$. In this case,
preheating ends before it is fully developed. Therefore, topological
defects are not formed, in this criterion, if 
$\eta \gtrsim 6.7 \times 10^{16} {\rm GeV}$. In fact, we find no
topological defects for $\eta=10^{17} {\rm GeV}$. However, this is not
a good criterion as we will see below. 

Actually, the amplitude of the fluctuations becomes a little smaller
than that of the homogeneous mode at most even if the breaking scale
is lower ($x^2 \simeq 0.25$ while $\langle \delta \varphi^2 \rangle
\simeq 0.2$), and they cannot affect the dynamics of the field wholly. 
This fact was also seen in the previous work using the Hartree
approximations which contain the forward scatterings only \cite{KK2}. 
Therefore, we conclude that the symmetry cannot be restored only
through the parametric resonance. However, after the preheating
ends, the mode-mixing (rescattering) of the fluctuations becomes
important for determining whether or not the topological defects are 
produced.

Now we are going further into the detail of the rescattering. At the
end of preheating, the occupation numbers of the fluctuations are very 
large only in the resonance band. The typical time scale for the
rescattering is estimated as follows \cite{KLS3}. The 
cross section for a scattering is naively given by   
$\sigma_1 \sim \lambda^2 / E^2$. Here $E \sim \sqrt{\lambda}\Phi$ is
the typical energy (momentum) of the produced particle. However, since
a huge amount of particles are present after preheating, the cross
section is much larger because of the Bose enhancement. Therefore the
phase space distribution function (the number density in $k$-mode)
$n_k \sim n_{\varphi}/E^3$ should be multiplied to $\sigma_1$. After
all, the scattering rate becomes 
\begin{equation}
  \label{gamma}
    \Gamma \sim \sigma_1 n_k n_{\varphi} \sim \sqrt{\lambda}\Phi.
\end{equation}
Using Eqs.(\ref{re-sf}) and (\ref{sf}), we get a typical time scale
for the
rescattering $\tau_{scat} \sim O(1)$. Then the occupation number at
higher momenta will be significant within $\Delta\tau \sim 30$. 
Notice that the effective rescattering is essentially the same
mechanism as the parametric resonance in the sense that these are
stimulated effects. The process 
$\delta\varphi(k_{res}) + \delta\varphi(k_{res}) \rightarrow 
\delta\varphi(k_{low}) + \delta\varphi(k_{high})$ first occurs. 
Typically, $k_{low}=0$ because of the Bose enhancement, since the
occupation number with $k=0$ (homogeneous mode) is large. In this
picture some major processes are as follows:
\begin{eqnarray}
    & & \delta\varphi(k_{res})  +  \delta\varphi(k_{res}) 
        \rightarrow   
        \varphi_0   +  \delta\varphi(2k_{res}),
        \nonumber \\
    & & \delta\varphi(2k_{res})  +  \varphi_0
        \rightarrow  
        \delta\varphi(k_{res})  +  \delta\varphi(k_{res}),
        \nonumber \\
    & & \delta\varphi(2k_{res})  +  \delta\varphi(2k_{res}) 
        \rightarrow   
        \delta\varphi(k_{res})  +  \delta\varphi(3k_{res}),
        \nonumber \\
    & & \delta\varphi(3k_{res})  +  \varphi_0
        \rightarrow   
        \delta\varphi(k_{res})  +  \delta\varphi(2k_{res}).
\end{eqnarray}
These are efficient because the produced fluctuations of mode $k$ 
(the first terms of the RHS's) have
already large occupation numbers. However, these processes
are the zero-sum game. If high momentum fluctuations are produced, 
the homogeneous mode is also created and cannot decay further.

There are also processes which do not feel the bose enhancement in the 
first place such as:
\begin{eqnarray}
    & & \delta\varphi(k_{res})  +  \varphi_0
        \rightarrow  
        \delta\varphi(k_1)  +  \delta\varphi(k_{res}-k_1),
        \nonumber \\
    & & \delta\varphi(2k_{res})  +  \varphi_0
        \rightarrow   
        \delta\varphi(k_2)  +  \delta\varphi(2k_{res}-k_2).
\end{eqnarray}
The field is knocked off from the homogeneous mode because of these
processes which make the spectrum smooth. Combining the above two
types of rescatterings, the occupation numbers of the mode outside 
the resonance band grow large enough for the zero mode to decay into
all the modes. The decay of the homogeneous mode completes in such a
complicated way, and the topological defects will be formed through
the Kibble-like mechanism later. Therefore, the rescattering timescale
for the defect formation is considerably longer than that given in
Eq.(\ref{gamma}). 

Now let us go on to the results of the rescattering effects. As we can
see in Fig.~\ref{fig-spec-1}, the fluctuations in the resonance band
develops in the first place, and later it moves to both higher and
lower momentum modes by the rescattering. The shape of the spectrum
becomes very broad and smooth, and further decays of the homogeneous
mode into all the modes can be seen because all the amplitudes grow
large. We show the case for $\eta=3\times 10^{16} {\rm GeV}$ in 
Figs.~\ref{fig-fl-2} and ~\ref{fig-spec-2}. Although the amplitude of
fluctuations becomes as large as that of $\eta=10^{16} {\rm GeV}$,
both higher modes and lower modes do not develop so much in comparison 
with Fig.~\ref{fig-spec-1}. Therefore, it is not enough time for
the rescattering to produce the topological defects when 
$\eta=3\times 10^{16} {\rm GeV}$. Actually, we find no topological
defects at the end of the calculation for this breaking scale.
Consequently, it is necessary to have a somewhat long stage
of the rescattering in order to produce the defects. We find that
the topological defects forms for  
$\eta \lesssim 2\times 10^{16} {\rm GeV}$ at most and can thus
conclude that duration of rescattering needs $\tau \gtrsim 200$.   

Here we investigate cases for $\eta=10^{16} {\rm GeV}$ and 
$\eta=2\times 10^{16} {\rm GeV}$ in some extent. In the latter case,
there are only a few strings in the horizon size (the whole region
at the end of calculation), as seen in Fig.~\ref{fig-r-def}. 
The identification of the defects is done by observing the winding at
the space point on the lattice as is seen in Fig.~\ref{fig-phase}.
Since half of strings are in fact anti-strings and each pair of the
string and anti-string is very close to each other, it 
will annihilate and disappear very soon, which means that the case of
$\eta=2\times 10^{16} {\rm GeV}$ may be harmless for cosmological
history. Actually, the numbers of defects at the end of calculations
are 2, 4, 2, and 16 for four runs. On the other hand, for the case of
$\eta=10^{16} {\rm GeV}$, even though the number of strings in the 
horizon decreases as time goes on, they still exist in a certain
amount (see Fig.~\ref{fig-p1-scale}). We expect that it will settle
down to some kind of a scaling solution later. The evidence that
topological defects will survive annihilations is that the
average number of strings at the end time of calculation
($\tau=2000$) is a little larger and the positions of strings and
anti-strings are not always pair-like, which is seen in 
Fig.~\ref{fig-p15-def} (d) as opposed for 
$\eta=2\times 10^{16} {\rm GeV}$ in Fig.~\ref{fig-r-def}.

%%%%%%%%%%%%%%%%%%%%%%%%%%%%%%%%%%%%%%%%%%%%
\section{Conclusions}
%%%%%%%%%%%%%%%%%%%%%%%%%%%%%%%%%%%%%%%%%%%%
We have considered the formation of topological defects after
inflation. To this end, we have integrated the equation which
describes the evolution of a complex scalar field with the potential 
$V(\Phi)=\lambda(|\Phi|^2-\eta^2)^2/2$ on the two-dimensional 
lattice with a slab symmetry in the $z$-direction in
order to include both parametric resonance 
effects and rescatterings of fluctuations. We have found that
fluctuations produced during preheating do not lead to the defect
formation even if their amplitudes become the same order of
magnitude as that of the homogeneous mode at the end of
preheating. We also have found that the effects of the rescattering of 
the fluctuations are essential and a rather long duration of the
rescattering is necessary for the topological defects to be formed
as suggested by Refs.~\cite{Kofman1,Kofman2}. Fluctuations in
the resonance mode scatter off each other to make fluctuations with
higher and lower momenta. They also knock the field off from the zero
mode to broaden and smooth the spectrum, which 
in turn stimulates further decays of the homogeneous mode into all the
modes. Therefore, the topological defects are formed in a similar way
as the Kibble mechanism. These processes take a lot of time so that
it is necessary for the field not to fall into the minimum of the
potential too soon after preheating. This constrains the value of the
breaking scale. From our calculations, we conclude that the
topological defects are not formed if 
$\eta \gtrsim (2 - 3) \times 10^{16} {\rm GeV}$.  

The constraint on the breaking scale is higher than that of our
previous result in Ref.~\cite{KK2} where the Hartree approximation is
adopted. In 
both cases, the amplitude of fluctuations at the end of preheating is
not larger than that of the homogeneous mode (compare 
Figs.~\ref{fig-fl-1} with \ref{fig-fl-3}). 
There are some differences between the
calculations with the Hartree approximation and the lattice simulation
during preheating. The fluctuations in the $y$-direction have a
similar shape, but the amplitude obtained with the use of
the Hatree approximation is a little smaller. The fluctuation in 
$x$-direction in the calculation with the Hatree approximation does
not grow until a very late 
time, which differs completely from the lattice one. The reason is
that most of the rescattering effects are neglected in the Hartree
approximation while the rescattering is very efficient for producing
fluctuations in both $x$- and $y$-directions in the lattice
calculations. Therefore, it is not
surprising that the critical breaking scale for the formation of
topological defects becomes higher in the lattice simulations.

{\it Note added}.
After submission of this paper, we noticed a paper,
Ref.~\cite{TKKL}, in which numerical calculations were done in three
dimensions, and similar results were reported. We agree with the 
authors of Ref.~\cite{TKKL} that rescatterings are less effective in two
dimensions to some extent. However, there is little difference in the
resonance effects. Actually, $\mu \simeq 0.141$ in ours [see the
footnote below Eq.(\ref{tau-ph})] while $\simeq 0.147$ in
Ref.~\cite{TKKL}. Moreover, the maximum value of fluctuations is
almost the same. The advantage of two-dimensional 
simulations is that the box size can be taken as large as the horizon
size at the end of calculations while the lattice size can be taken
small enough to identify defects. Therefore, we can estimate the
average numbers of defects per horizon size.

%\section*{Acknowledgment}

\end{document}